\documentclass[aps,preprint]{revtex4}

\def\.{\!\cdot\!}
\def\:{\cdots}
\def\[{\left[}
\def\]{\right]}
\def\({\left(}
\def\){\right)}

\def\bi{\begin{itemize}}

\def\ei{\end{itemize}}
\def\be{\begin{eqnarray}}
\def\ee{\end{eqnarray}}
\def\bn{\begin{enumerate}}
\def\en{\end{enumerate}}
\def\h{{1\over 2}}

\def\nn{\nonumber}

\def\r2{\sqrt{2}}

\def\e{\epsilon}
\def\L{{\cal L}}
\def\d{\delta}
\def\rr2{{1\over\sqrt{2}}}
\def\th{\theta}

\begin{document}
\title{Neutrino 2-3 Symmetry and Inverted Hierarchy}
\author{C.S. Lam}
\address{Department of Physics and Astronomy, University of British Columbia,  Vancouver, BC, Canada V6T 1Z1 \\
and\\
Theory Group, TRIUMF, Vancouver, BC, Canada V6T 2A3\\
\email: Lam@physics.ubc.ca}

\begin{abstract}
Neutrino oscillation data indicate the presence of a 2-3 symmetry for the left-handed neutrinos.
We show that this symmetry between the second and third generations
cannot be extended to include the charged leptons, no matter what basis we use. We
point out that if this symmetry is also independently valid 
for the right-handed neutrinos, then the active neutrino spectrum is inverted,
with $m_3=0$. This conclusion remains valid even when the left-handed 2-3 symmetry is broken
by a non-maximal atmospheric mixing angle $\theta_{23}$, or a non-zero reactor angle $\theta_{13}$.  As previously 
shown by Mohapatra, Nasri, and Yu, such a symmetry also gives rise to interesting consequences on the leptogenesis
asymmetry parameter $\epsilon_1$.
\end{abstract}
\maketitle

\section{2-3 symmetry for left-handed leptons}
The leptonic mass Lagrangian can be written symbolically as
\be
\L=-\bar L_e M_e R_e-L_\nu^T M_\nu L_\nu,\ee
where $L_e, L_\nu$ are the left-handed fields for the charged leptons and neutrinos, and $R_e, R_\nu$ are the
corresponding right-handed fields.
$M_e$ is the $3\times 3$ charged lepton mass matrix, and $M_\nu$ is the symmetric mass matrix for 
left-handed Majorana neutrinos. In type-I seesaw, this Majorana mass is related to  
the Dirac mass  $-\bar L_\nu M_D R_\nu$ and the right-handed Majorana mass  $-R_\nu^TM_RR_\nu$ 
by
\be
M_\nu=M_D^TM_R^{-1}M_D.\ee

In the basis where $M_e={\rm diag}(m_e,m_\mu,m_\tau)$ is diagonal, $M_\nu$ can be diagonalized 
by the unitary MNS mixing matrix \cite{MNS} $U$, $M_\nu=U^TM_\nu^dU$, with $M_\nu^d={\rm diag}(m_1,m_2,m_3)$
being the (generally complex) neutrino mass parameters. Data are consistent \cite{ALTARELLI} with having
the atmospheric mixing angle $\theta_{23}$ maximal, and the reactor angle $\theta_{13}$ zero.
In that case, $U$ can be parameterized as
\be U={1\over \r2}\pmatrix{\r2 c&\r2 s&0\cr -s&c&1\cr -s&c&-1\cr},\ee
where $s=\sin\theta_{12}$, $c=\cos\theta_{12}$, and $\theta_{12}$ is the solar mixing angle.

The resulting neutrino mass matrix 
\be 
M_\nu=U^TM_\nu^dU=
\pmatrix{\nu_{11}&\nu_{12}&\nu_{12}\cr \nu_{12}&\nu_{22}&\nu_{23}\cr \nu_{12}&\nu_{23}&\nu_{22}\cr},\ee
with
\be
\nu_{11}&=&c^2m_1+s^2m_2\nn\\
\nu_{12}&=&cs(m_2-m_1)/\r2\nn\\
\nu_{22}&=&\h(s^2m_1+c^2m_2+m_3)\nn\\
\nu_{23}&=&\h(s^2m_1+c^2m_2-m_3),\ee
is invariant under the simultaneous interchanges of the second and third columns, 
together with the second and third rows. We shall refer to this symmetry as the 2-3 symmetry
for left-handed neutrinos.

Conversely, the invariance of $\L$ under a permutation of the second and third generations $L_\nu$,
via the matrix
\be
P=\pmatrix{1&0&0\cr 0&0&1\cr 0&1&0\cr},\ee
will result in an identity $L_\nu^TPM_\nu PL_\nu=L_\nu^TM_\nu L_\nu$, or 
equivalently $PM_\nu P=M_\nu$. The most general symmetric
matrix satisfying this constraint is of the form given by (4). Moreover, a matrix of this form is known to lead
to a maximum $\theta_{23}$ and a zero $\theta_{13}$ \cite{LAM,OTHERS}.

Since left-handed neutrinos have a 2-3 symmetry, one might wonder whether the left-handed charged leptons 
also have the same 
symmetry. In the basis where the charged leptons are diagonal, clearly they do not, because $m_\mu\not= m_\tau$.
Actually, a 2-3 symmetry cannot be simultaneously valid for the left-handed charged leptons and 
the left-handed neutrinos, {\it no matter what basis we choose}.
This can be seen as follows.

If both $L_\nu$ and $L_e$ are 2-3 symmetric, then $PM_\nu =M_\nu P$, and $PH_e =H_eP$, where $H_e=M_eM_e^\dagger$.
Since they commute, there must be a unitary matrix $V_\nu$ that can diagonalize $P$ and $M_\nu$ simultaneously, 
and another unitary matrix $V_e$ that can diagonalize $P$ and $H_e$ simultaneously. Now the matrix $S$
below diagonalizes $P$,
\be
S^\dagger P S=\pmatrix{1&0&0&\cr 0&1&0&\cr 0&0&-1&\cr}\quad{\rm for}\quad S=\pmatrix{1&0&0\cr 0&1/\r2&1/\r2\cr 0&1/\r2&-1/\r2\cr},\ee
but since $P$ has two degenerate eigenvalues $+1$, this diagonalization is not unique.
The most general unitary matrix to diagonalize $P$ is of the form $V=SW$, with
\be W=\pmatrix{w&0\cr 0&e^{i\phi}},\ee
and $w$ any $2\times 2$ unitary matrix. The neutrino mixing matrix $U=V_e^\dagger V_\nu=W_e^\dagger W_\nu$
is therefore block diagonal, implying no mixing of the third-generation neutrino with the other two 
neutrino generations. This
is manifestly false, hence the 2-3 symmetry cannot be simultaneously true for $L_e$ and $L_\nu$
in any basis. 

We do not understand why the 2-3 symmetry is (at least approximately) true for the left-handed
neutrinos but not for the left-handed charged leptons, though there are models that can describe
it \cite{OTHERS},
but data indicate that to be so. From now on
we will concentrate solely on the neutrino symmetry.

\section{2-3 symmetry for the right-handed neutrinos}
Given a mass matrix $M_\nu$ in (4) and (5), the seesaw formula (2) is insufficient to determine
both the  Dirac mass matrix $M_D$ and the right-handed Majorana mass matrix $M_R$. 
$M_R$, being symmetric,
contains 6 complex parameters, but $M_D$, not having any symmetry,
 generally contains 9. Since $M_D^TM_R^{-1}M_D$ is symmetric, 
the seesaw formula contains 6 constraints, thereby leaving 9 complex parameters 
in $M_D$ and $M_R$ undetermined. If we want to use neutrino oscillation to probe the dynamics beyond
the Standard Model, we need to know those parameters. In any case, 
some combinations of these parameters are actually required to calculate the asymmetry parameter $\e_1$
 in leptogenesis. 

One might be able to reduce the number of undetermined parameters if the
right-handed neutrinos obey some symmetry.
Since the left-handed neutrinos exhibit a 2-3 symmetry, we shall assume the right-handed
neutrinos to possess a 2-3 symmetry as well. 
This includes the possibilities of having the right-handed neutrinos possess a 1-2 or a 1-3 symmetry instead, 
because  a relabeling of generations of the right-handed neutrinos will convert those symmetries into a 2-3 symmetry.

The right-handed 2-3 symmetry might be coupled and linked to the left-handed 2-3 symmetry, or it might be 
an independent symmetry 
not linked to the left-handed one. Let us look at these two cases separately.

In the first case, $\bar L_\nu M_{D}R_\nu=\bar L_\nu PM_{D}PR_\nu$ and $R_\nu^T M_RR_\nu=R_\nu^TPM_RPR_\nu$,
so the constraints on the mass matrices are $M_D=PM_{D}P$, and $M_R=PM_RP$. The constraint on $M_R$ is
identical to the constraint on $M_\nu$, so $M_R$ contains 6 complex parameters. $M_D$, not being 
symmetric, contains 7 complex parameters. The number of undetermined parameters is now reduced from
9 to 7.

As shown by Mohapatra and Nasri \cite{MN1}, and also by Grimus and Lavoura \cite{GL},
the leptogenesis asymmetry parameter $\e_1$ is then
proportional to $\Delta m_\odot^2$, rather than the more generic $\Delta m_{atm}^2$ \cite{BUCH}.

If the right-handed 2-3 symmetry is independent of the left-handed 2-3 symmetry, then
the constraint on $M_R$ remains the same, but the
constraint on $M_D$ becomes stronger. In that case,  
$\bar L_\nu M_{D}R_\nu=\bar L_\nu PM_{D}R_\nu=\bar L_\nu M_{D}PR_\nu$,
hence
$M_D=PM_D=M_DP$. In other words, the second and third columns of $M_D$ must be identical,
and so must be the second and third rows. Thus $M_D$ is of the form
\be
M_D=\pmatrix{a&b&b\cr d&c&c\cr d&c&c\cr},\ee
which is
specified by  4 complex parameters. In this case
the number of unknown parameters is further reduced from 7 to 4.

The conclusion of Ref.~\cite{MN1} and \cite{GL} on $\e_1$ remains valid, but there are now additional consequences which
we shall discuss in the rest of this article.

Since $M_D$ has two identical columns, its determinant is zero. Hence
$M_\nu$ given by (2) also 
has a zero determinant, and thus a zero eigenvalue.
This eigenvalue is $m_3$, as can be seen as follows.
Compute $S^TM_\nu S$, where $M_\nu$ is the matrix in (4). It yields a block diagonal matrix, with
the 13, 23,  31, and 32 matrix elements zero,
and the 33 matrix element equal to $\nu_{22}-\nu_{23}$. Using (5), this difference is equal to $m_3$. 
Now compute $S^TM_\nu S$ again, this time using $M_\nu$ given by (2). With $M_R$ 
2-3 symmetric and $M_D$ given by (9), a straight forward
calculation shows that the resulting 33 matrix element is zero. Hence $m_3=0$.

With $m_3=0$, the magnitudes of the remaining neutrino masses are determined by the atmospheric and solar
mass gaps to be
\be
|m_1|&=&\sqrt{\Delta m_{atm}^2}\simeq 52\ {\rm meV}\nn\\
|m_2|&=&\sqrt{\Delta m_{atm}^2+\Delta m_\odot^2}\simeq 53\ {\rm meV}.\ee
The sum $\sum_{i=1}^2|m_i|$, just over 0.1 eV, is comfortably below the upper bound of 0.47 eV placed by
astrophysical data, including Ly$\alpha$ \cite{ASTRO}.
The effective neutrino mass measured in tritium $\beta$-decay is 
\be
m_{\nu_e}=\sqrt{\sum_i|U_{ei}^2m_i^2|}=\sqrt{c^2|m_1|^2+s^2|m_2|^2},\ee
which lies between $|m_1|$ and $|m_2|$.
This is to be compared with the 2.2 eV upper bound from experiments \cite{TRITIUM}. Unfortunately, this number
is too low to be reached by KATRIN for verification.
Finally, the effective mass for the neutrinoless double $\beta$-decay is
\be
m_{ee}&=&|\sum_{i=1}^3U_{ei}^2m_i|=|c^2 m_1+s^2 m_2|
=\sqrt{(c^2|m_1|+s^2|m_2|)^2-|m_1m_2|\sin^22\theta_\odot\sin^2{\phi_{12}\over 2}}\nn\\
&\simeq& |m_1|\sqrt{1-0.84
\sin^2{\phi_{12}\over 2}},\ee
where $\phi_{12}$ is the relative Majorana phase angle between $m_1$ and $m_2$. The current
upper bound from experiment is about 0.3 eV, with a factor of 3 uncertainty from the nuclear matrix elements \cite{ZB}.

\section{Breaking the left-handed 2-3 symmetry}
If $\theta_{23}$ is not exactly maximal, or $\theta_{13}\not=0$, then the 2-3 symmetry for the left-handed
neutrino is broken. Nevertheless, present experiments indicate that the breaking has to be small.

We can quantify the breaking in the following way.
Under a 2-3 permutation of the left-hand neutrino field, $L_{\nu}\to PL_{\nu}$, the mass Lagrangian $\L$ can be decomposed into 
the sum of an even part and an odd part: $\L=\L^e+\L^o$, with $\L^e\to \L^e$
and $\L^o\to-\L^o$. This induces
a decomposition of the mass matrix $M_\nu=M_\nu^e+M_\nu^o$, with the constraints
$M_\nu^e=PM_\nu^eP$ and $M_\nu^o=-PM_\nu^oP$. Thus $M_\nu^e$ is of the form given by (4), but $M_\nu^o$ is of the form 
\be
M_\nu^o=\pmatrix{0&\nu'_{12}&-\nu'_{12}\cr \nu'_{12}&\nu'_{22}&0\cr  -\nu'_{12}&0&-\nu'_{22}\cr}.\ee
To calculate $\nu'_{ij}$, 
we assume the atmospheric mixing angle to be maximal, and
the reactor angle $\theta_{13}$ to be small but not necessarily zero. Then instead of (3), the MNS matrix 
to $O(\th)$ is given by
\be
U={1\over \r2}\pmatrix{\r2 c&\r2 s&\theta\cr -s+c\th^*&c+s\th^*&1\cr -s+c\th^*&c+s\th^*&-1\cr},\ee
 where $\th\equiv\sin\th_{13}e^{i\d}$ and $\d$ is the CP phase. 
By calculating $M_\nu=U^TM_\nu^dU$ up to $O(\th)$, we obtain the even part $M_\nu^e$ to be given by
(4) and (5), and the odd part $M_\nu^o$ to be  given by (13), with
\be
\nu'_{12}&=&(m_3-c^2m_1-s^2m_2)\th/\r2,\nn\\
\nu'_{22}&=&cs(m_1-m_2)\th.\ee
This is small because of the CHOOZ bound \cite{CHOOZ}.

How does this breaking of the left-handed neutrino affect 
the 2-3 symmetry of the right-handed neutrino? If its 2-3 symmetry is coupled and linked to the 2-3
symmetry of the left-handed neutrino, we expect a breaking of the right-handed
symmetry as well. Since there are no data to guide us how the right-handed
neutrinos behave, there are many uncertainties connected with this scenario.

If the 2-3 symmetry of the right-handed neutrino is independent 
and unlinked to the 2-3 symmetry of the left-handed
neutrino, then there is no need to break the right-handed symmetry when the left-handed symmetry is broken. 
In that case, $M_R$ remains bound by the constraint $M_R=PM_RP$, but $M_D$ is bound only by the constraint
$M_D=M_DP$ and {\it not} the constraint $M_D=PM_D$, as the left-handed 2-3 symmetry is broken. Nevertheless,
since the second and third columns of $M_D$ are still identical, its determinant is still zero. The conclusion
of having $m_3=0$ therefore remain unchanged. However, since $\theta_{13}\not=0$, the previous estimate
of $m_{\nu_e}$ should be multiplied by a factor $\cos\theta_{13}$, and that of $m_{ee}$ should be
multiplied by a factor $\cos^2\theta_{13}$.

The consequence for the leptogenesis asymmetry parameter $\e_1$ in this scenario is discussed in Ref.~\cite{MN2}.
$\e_1$ is now proportional to a linear combination of $\Delta m_\odot^2$ and $|\th|^2$.

In conclusion, we found that in the presence of a 2-3 symmetry for the right-handed neutrinos, the 
spectrum for the active
neutrinos  is inverted, with $m_3=0$. This conclusion remains valid whether the left-handed 2-3 symmetry is broken or not.

This research is supported by the Natural Sciences and Engineering Research Council of Canada.



\begin{thebibliography}{9}

\bibitem{MNS} Z. Maki, M. Nakagawa, and S. Sakata, Prog. Theo. Phys. 28
(1962) 870.
\bibitem{ALTARELLI} For a review and references to the original literature, see G. Altarelli, hep-ph/0410101.
\bibitem{LAM} C.S. Lam, Phys. Lett. B507 (2001) 214.
\bibitem{OTHERS} T. Kitabayashi and M. Yasue, Phys. Rev. D67 (2003) 015006; W. Grimus
and L. Lavoura, Phys. Lett. B572 (2003) 189; J. Phys. G30 (2004) 73; hep-ph/0408123;  Y. Koide, H. Nishiura, K. Matsuda, T. Kikuchi, 
and T. Fukuyama, Phys. Rev. D66 (2002) 093006;
Y. Koide, Phys. Rev. D69 (2004) 093001;
R. N. Mohapatra, SLAC Summer Inst. lecture; http://www-conf.slac.stanford.edu/ssi/2004;
hep-ph/0408187; JHEP, 0410, 027 (2004); W. Grimus, A. S.Joshipura, S. Kaneko, L. Lavoura,
H. Sawanaka, M. Tanimoto, hep-ph/0408123; X.-G. He and A. Zee, Phys. Lett. B560 (2003) 87.
\bibitem{MN1} R.N. Mohapatra and S. Nasri,  Phys. Rev. D71 (2005) 033001.
\bibitem{GL} W. Grimus and L. Lavoura, J. Phys. G30 (2004) 1073.
\bibitem{BUCH} W. Buchmuller, M. Plumacher and P. di Bari, Phys. Lett. B547 (2002) 128.
\bibitem{ASTRO} U. Seljak et al., astro-ph/0407372; G.L. Fogli et al., Phys. Rev. D70 (2004) 113003.
\bibitem{TRITIUM} K. Eitel, "Direct Neutrino Mass Experiments," in Neutrino 2004:
21st International Conference on Neutrino Physics and Astrophysics (Paris, France, 2004).
\bibitem{ZB}  H.V. Klapdor-Kleingrothaus, Eur. Phys. J. A12 (2001) 147.
\bibitem{CHOOZ} M. Apollonio et al., Phys. Lett. B420 (1998) 397.
\bibitem{MN2} R.N. Mohapatra, S. Nasri, and H. Yu, hep-ph/0502026.

\end{thebibliography}
\end{document}